\documentclass{aa} 
\usepackage{graphicx}
\usepackage{txfonts}
\newcommand{\etal}{{et al. }}
\newcommand{\figwidth}{84mm}
\newcommand{\be}{\begin{equation}}
\newcommand{\ee}{\end{equation}}
\begin{document}
   \title{Site testing in summer at Dome C, Antarctica}

   \subtitle{}

   \author{E. Aristidi\inst{1}
          \and
    A. Agabi\inst{1}
         \and
  E. Fossat\inst{1} \and
  M. Azouit\inst{1} \and
   F. Martin\inst{1} \and
 T. Sadibekova\inst{1}\and
  T. Travouillon\inst{2}\and
J. Vernin\inst{1}\and
  A. Ziad\inst{1}
           }

   \offprints{E. Aristidi, \email{aristidi@unice.fr}}

   \institute{Laboratoire Universitaire d'Astrophysique de Nice,
              Universit\'e de Nice Sophia Antipolis, Parc Valrose,
          06108 Nice Cedex 2, France
      \and School of Physics,University of New South Wales, Sydney,
              NSW 2052, Australia       }

   \date{Received: May 2005 / Accepted:}

   \abstract{We present summer site testing results based on DIMM data obtained at Dome C, Antarctica. These data have been collected on the bright star Canopus during two 3-months summer campaigns in 2003-2004 and 2004-2005. We performed continuous monitoring of the seeing and the isoplanatic angle in the visible. We found a median seeing of 0.54~\arcsec and a median isoplanatic angle of 6.8~\arcsec. The seeing appears to have a deep minimum around 0.4 \arcsec almost every day in late afternoon.

   \keywords{Site Testing, Antarctica
                               }
   }

   \maketitle
%

\section{Introduction}
The French (IPEV) and Italian (ENEA) polar Institutes are
constructing the Concordia base on the Dome C site of the Antarctic
plateau (75°S, 123°E), at an elevation of 3250~m, that corresponds,
given the cold air temperatures, to an air pressure encountered
around 3800~m at more standard latitudes. The Concordia construction
is now completed, the first winterover has started in 2005.
Astronomy is obviously near the top of the list of the scientific
programmes that will benefit of this unique site: the extremely cold
and dry air is complemented by very low winds, both at ground level
and at higher altitude, so that an exceptionally good seeing is
expected.

In the late 90's a site testing program based on balloon-borne
microthermal sensors has been conducted by J. Vernin and R. Marks at
South Pole (Marks \etal \cite{marks}). As katabatic winds are
present at South Pole, these authors found poor seeing from the
ground (recent measurements by Travouillon \etal\ (\cite{trav})
confirmed a value of 1.7\arcsec\ in the visible range). An amazing
result is that the seeing drops down to 0.3\arcsec\ at 200~m height
above the surface, i.e. at an altitude of 3050~m above the sea
level. Therefore Dome C with its 3250~m altitude and located in a
low wind area, appeared as an excellent candidate for astronomy.

These promising qualities encouraged our group to initiate a
detailed analysis of the astronomical site properties. In 1995, a
Franco-Italian group directed by J. Vernin made a pioneering
prospective campaign at Dome C and launched a few meteorological
balloons. A systematic site testing program was then initiated,
under the name of Concordiastro, first funded by IPEV in 2000.
Proposed site testing was based upon two kinds of measurements.
First, a monitoring of the turbulence parameters in the visible
(seeing $r_0$, isoplanatic angle $\theta_0$, outer scale ${\cal
L}_0$ and coherence time $\tau_0$) with a GSM experiment (Ziad \etal
\cite{ziad}) specially designed to work in polar winter conditions.
In addition to this monitoring, it was proposed to launch balloons
equipped with microthermal sensors to measure the vertical profile
of the {\bf refractive index structure constant} $C_n^2(h)$
(Barletti \etal \cite{barletti}). 50 to 60 balloons were foreseen to
be regularly launched during the polar winter.

On-site campaigns began in summer 2000-2001 and were performed every
year until 2004-2005 with a double aim. It was first necessary to
test the behaviour of all instruments in the ``intermediate" cold
temperature of the summer season, ranging between -20 and
-50$^\circ$C. This first step was also used to anticipate the more
difficult winter conditions, with -50 to -80$^\circ$C, for being
technically confident with the first winterover equipment. On the
other hand, the summertime sky quality is interesting in itself as
solar observations were started in 1979-1980 at the South Pole. At
Dome C the long uninterrupted sequences of coronal sky (far longer
than at South Pole) and the expected occurrence of excellent seeing
make it a very promising site for high resolution solar imaging and
specially solar coronography.

So far 6 summer campaigns have been performed totalling 80 man-week
of presence on the site. The first winterover has also begun this
year. 197 meteorological balloons have been successfully launched,
corresponding results have been published in Aristidi \etal\
(\cite{windpaper}). Major conclusions are that the wind speed
profiles in Dome C appears as the most stable among all the
astronomical sites ever tested, and that the major part of the
atmospheric turbulence is probably generated in the first 100~m
above the snow surface, where the temperature gradients are the
steepest (around 0.1 $^\circ$C/m).

In this paper we present the results of daytime turbulence
measurements (seeing and isoplanatic angle) made with various
techniques. The paper is organised as follows: in Sect.~2 we briefly
review the theory of the turbulence parameters we measured. Section
3 presents the instrumental setup at Dome~C. Section 4 describes the
observations, the various calibration procedures and the online and
offline data processing. {\bf The results of the monitoring are in
Sect.~5. A final discussion is presented in Sect.~6 and an appendix
on error analysis ends the paper.}


\section{Theory}
\subsection{Seeing}
Atmospheric turbulence is responsible for the degradation of image
resolution when observing astronomical objects. The full width at
{\bf half maximum} (FWHM) of the long-exposure point-spread function
broadens to a value $\epsilon$ called {\em seeing}, usually
expressed in arcseconds, this parameter represents the angular
resolution of images for given atmospheric conditions. In the
visible $\epsilon$ is around 1\arcsec\ for standard sites.

Fried (\cite{fried}) introduced a so-called parameter $r_0$ that can
be regarded as the diameter of a telescope whose Airy disc has the
same size than the seeing. He derived the following relation \be
\epsilon=0.98\; \frac{\lambda}{r_0} \ee

The seeing is one of the most important parameters that describes
atmospherical turbulence. Seeing monitors have been installed in
major observatories such as ESO Paranal and produce constant data
that are used to optimize the observations. Seeing estimation can be
made by various methods (Vernin \& Munoz \cite{verninmunoz}); seeing
monitors that allows continuous measurements are traditionally based
on differential image motion such as the DIMM (Differential Image
Motion Monitor) used at Dome C. It is extensively described in the
literature (Sarazin \& Roddier \cite{roddiersarazin}, Vernin \&
Munoz \cite{verninmunoz}, Tokovinin \cite{toko}) and has become very
popular because of its simplicity.

A DIMM is a telescope with an entrance pupil made of 2
diffraction-limited circular sub-apertures of diameter $D<r_0$,
separated by a distance $B>D$. A tilt is given to the light
propagating through one of the two apertures to produce twin images
that move according to the turbulence. Fried parameter is computed
from longitudinal ($\sigma_l^2$) and transversal ($\sigma_t^2$)
variances of the image motions {\bf using equations 5 and 8 of
Tokovinin (\cite{toko})}.

\subsection{Isoplanatic angle}
The isoplanatic angle is a fundamental parameter for adaptive optics
(AO). It is the correlation angle of the turbulence, i.e. the
maximum angular distance between two point-sources affected by the
same wavefront distortions. In AO systems, these distortions are
usually estimated on a nearby bright reference star. This reference
star must be in the isoplanatic domain, which in most cases reduces
dramatically the observable piece of sky.

As for the seeing, the isoplanatic angle $\theta_0$ is a scalar
random variable, usually expressed in arcseconds, resulting from an
integral over the $C_n^2$ profile (Ziad \etal \cite{ziad}, Avila
\etal \cite{avila}). Loos \& Hogge (\cite{loos}) proposed an
approximate estimation based on the scintillation of a point-source
star through a 10~cm pupil. Estimation is even better if one uses a
4~cm diameter central obstruction. As for the DIMM, this technique
allows a continuous monitoring of the isoplanatic angle as well as
the scintillation data. It is used routinely by the GSM instrument
(Ziad \etal  \cite{ziad}) for site qualification.


\section{Instrumentation}

\subsection{The Concordiastro observatory}
\begin{figure}
\includegraphics[width=\figwidth]{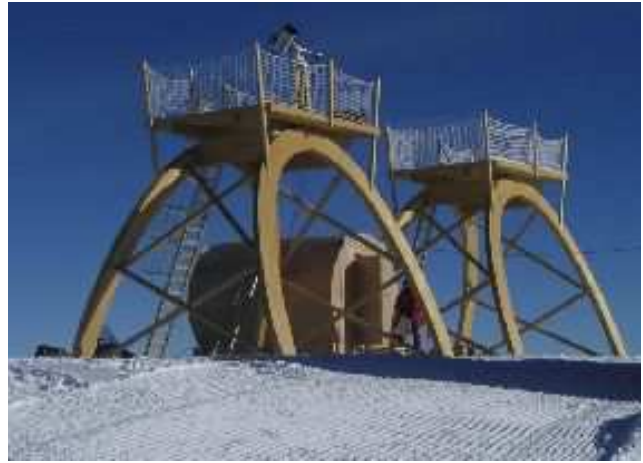}
\caption{Concordiastro observatory in November 2004. One can see the two platforms and the wooden igloo between. The DIMM is on the top of one platform.}
\label{fig:platform}
\end{figure}

The Concordiastro observatory is based on two wooden platforms
designed by J. Dubourg (Observatoire de la C\^ote d'Azur) and built
by the ``Ateliers Perrault Fr\`eres'', a factory of western France.
The design of these platforms recalls the first floor of the Eiffel
Tower (see Fig.~\ref{fig:platform}). These platforms are 5~m high
and equipped with massive supports for the telescopes. The height
has been chosen for site-testing purpose to avoid the surface layer
turbulence. They are located 300~m away of the Concordia station, in
South-West direction to avoid pollution (wind comes from South or
South-West most of the time (Aristidi \etal \cite{windpaper})). The
first one was erected in December 2002, the second one in  January
2004. All the installation is built onto a 2~m high pavement of
compacted snow for stability (the same kind of pavement that
supports the Concordia buildings).

Between the two platforms, a wooden container nicknamed ``igloo''
hosts the electronics and the control systems. It was installed at
the very end of the 2003-2004 summer campaign. It is cabled to the
telescopes, and a fiber optic link is foreseen for remote control
from Concordia towers during the winter.

Additional telescopes have been installed at 1.50~m above the ground
during the campaign 2004-2005 for estimating the surface layer
contribution in the seeing.

\subsection{Telescopes}
We use Schmidt-Cassegrain Celestron C11 telescopes (diameter 280~mm)
with a $\times 2$~Barlow lens (equivalent focal length 5600~mm).
Optical tubes have been rebuilt in INVAR to reduce thermal
dilatations (these dilatations cause defocus since they change the
distance between the primary and secondary mirrors). Several
technical improvements have been made on the primary mirror support,
and the grease of the focus system have been replaced by a
cold-resistant one (up to -90$^\circ$ C).

These telescopes are placed on equatorial mounts Astro-Physics~900.
Here again, some customization has been performed: grease was
changed and heating systems were placed in the motor carters.
Tracking has worked successfully during the entire polar summers.
The mounts are placed  on massive wooden feet fixed to the platform.
Polar alignment is made by Bigourdan's method on solar spots
(fortunately we were close to a solar maximum and finding spots had
never been a problem), then on Venus, and fine tuning was made on
Canopus itself during the observations.

\subsection{Cameras}
At the focus of all our telescopes we use a digital CCD camera (PCO
Pixelfly) connected to a PCI board via a high speed transfer cable.
Technical specifications are given in Table~\ref{table:camera}.
\begin{table}
\caption{\bf Technical specifications of the CCD camera}
\begin{tabular}{|l|l|}\hline
 Number of pixels: & 640 $\times$ 480\\
 Pixel size: & $9.9 \times 9.9 \mu$m\\
 Binning modes: & horizontal: 1,2 \\
 & vertical: 1,2, 4\\
 Dynamic range: & 12 bits\\
 Exposure time: & 10$\mu$s to 10 s\\
 Frame rate & 40 fps without binning\\
 & 76 fps in binning $2 \times 2$\\
 Maximum QE & 40\% at 350 and 500~nm\\
  Bandwith (FWHM) & 320 -- 630 nm\\
 ADU & 7 e$^-$/count\\
 Readout noise: & 16 e$^-$\\ \hline
\end{tabular}
\label{table:camera}
\end{table}
The camera was placed into an insulated and thermally controlled
box, insulation being inherited from spatial technology. Typical
temperature inside the box was around -15$^\circ $C, and over
$0^\circ$C on the CCD chip thanks to dissipation. The $\times
2$~Barlow lens was placed at the box entrance.

\section{Observations and data processing}
\subsection{Seeing measurements with the DIMM}

Dome C DIMM is based on a Celestron 11 telescope equipped with a 2
holes mask at its entrance pupil. Each sub-aperture has a diameter
of 6~cm and are separated with 20~cm. One is equipped with a glass
prism giving a small deviation angle (1 arcmin) to the incident
light. The other is simply closed with a glass parallel plate. The
size of the Airy disc is $\lambda/D=40\; \mu$m at the operating
wavelength (visible) that is compatible with Shannon sampling in
$2\times 2$ binning mode (effective pixel size is 20~$\mu$m. The
separation of the two images in the focal plane is 1.6~mm (80
pixels).

After different trials, we selected the star Canopus ($\alpha$~Car,
V=-0.7) for seeing monitoring. It is circumpolar at Dome C, with
zenithal angles $z$ ranging between 22$^\circ$ and 52$^\circ$. At
the end of December, Canopus and the Sun have 12~hour difference in
right ascension so that Canopus is at its maximum (resp. minimum)
elevation when the Sun is at its minimum (resp. maximum). The
angular distance between the two bodies remains around 100$^\circ$
during the whole summer season.

Three DIMM campaigns have been performed so far. The first one, in
2002-2003 (Aristidi et al., \cite{aaletter}, \cite{capri}) led to
seeing values that appeared since to be over-estimated. The
telescope used was black and heated by the Sun: we had strong local
turbulence that destroyed sometimes the Airy discs of the star
images into speckle patterns. {\bf We noticed evidence of this local
turbulence by {\em a posteriori} comparison between data taken
simultaneously from white and black telescopes.} The 2002-2003 data
we will not be taken into account in this paper.

{\bf The seeing values presented here have been collected during the
periods of Nov.~21, 2003-- Feb.~2, 2004 and Dec.~4, 2004-- Feb.~28,
2005. Data are also available beyond March 2005 but this paper deals
with summer conditions and we limited the data sample to the
daytime. Autumn and winter seeing will be discussed in forthcoming
papers. }

\subsubsection{Sky Background}
\begin{figure}
\includegraphics[width=\figwidth]{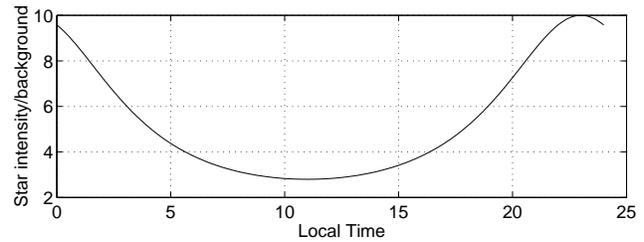}
\caption{Plot of the star image peak intensity {\bf $I_m$ to the sky background level model $\langle B(t)\rangle$. From images taken in the period Dec 1-15, 2003.}}
\label{fig:levelvstime}
\end{figure}

The sky background level is a strong limitation in daytime stellar
observations. {\bf We decided to quantify this background in the
first half of December 2003. From each image taken in that period we
measured the sky background $B(t)$ as a function of local time $t$.
We then performed a sinusoidal fit giving an empirical model for the
mean sky background $\langle B(t)\rangle$ as a function of local
time. We also measured and averaged the peak intensity of the star
images $I_m$. Figure~\ref{fig:levelvstime} plots the ratio
$I_m/\langle B(t)\rangle$ as a function of the local time}.
Background level is always between 10\% and 30\% of the stellar
flux, that is low enough to apply a threshold and still keep enough
stellar flux to make measurements. We were then able to perform, to
our knowledge, the first DIMM measurements ever in daytime, which
can be credited to the exceptional quality of the Dome C sky
appearing to be coronal a large fraction of the time.

\subsubsection{Exposure time}
The Fried's parameter must be ideally expressed for instantaneous
images. As a finite exposure time is used by the camera, there is an
exposure bias that must be removed. The technique, described by
Tokovinin (\cite{toko}), consists in making successive poses using
alternate exposure times $\tau$, $\tau/2$, $\tau$, $\tau/2$, \ldots
and performing a modified exponential extrapolation to attain
instantaneous values: \be \epsilon(0)=\epsilon(\tau)^{1.75}\;
\epsilon(\tau/2)^{-0.75} \ee where $\epsilon(x)$ is the seeing
estimated with exposure time $x$. We chose $\tau=10$~ms, that had
the double advantage of exploiting the entire CCD dynamics and to be
a standard for seeing monitors. {\bf We observed that the correction
depends on the turbulence conditions; it is close to zero when the
seeing is good and can grew up to 20\% when $\epsilon >
1.5$~\arcsec.  A few percent difference between transverse and
longitudinal seeing values has also been noticed ; it is a
well-known effect (Martin \cite{martin87}) related to the wind speed
and direction and to the exposure time}.

\subsubsection{Seeing estimations}
Times series were divided in 2~minutes intervals in which around
9000 short-exposure frames were acquired using the $2\times 2$
binning mode of the CCD. 2 minutes is a time interval large enough
to saturate the structure function of the motion of DIMM images. A
software was developed to perform real time data processing. Each
short exposure frame was flat-fielded to eliminate the background,
then the two stellar images were easily detected in two small
$20\times 20$ pixels windows and their photocentre coordinates
computed by means of a simple barycenter formula. Note that with the
flat-fielding, the effective illuminated pixels correspond roughly
to the surface of the Airy disc of the sub-apertures. Every two
minutes, the variances of longitudinal and transverse distances
between the two images were computed in units of pixel square, then
converted into arcsec using the scale calibration described below.
This leads to two independent estimates of the Fried parameter,
namely $r_{0l}$ (longitudinal) and $r_{0t}$ (transverse) which are
stored in a file. Then the two $20\times 20$ pixels windows are
moved so that their center is placed on the previous photocentres of
the two stars images for the following seeing estimation. This
allows a gain of time in the barycenter calculation, and to follow
the stars if they move in the field of view (guiding problems for
example).

Three corrections are then made in post-processing to obtain actual seeing values:
\begin{itemize}
\item Transversal and longitudinal seeings are computed and corrected from exposure time as described above.
\item As the seeing is a scalar parameter, both transverse and longitudinal estimations should give the same value. We kept only pairs verifying $
0.7<\epsilon_{t}/\epsilon_{l}<1.3$ (around 90\% of the data sample).
 Longitudinal and transverse values are then averaged.
\item {\bf Finally we made compensation from zenithal distance $z$ (Tokovinin \cite{toko})}.
\end{itemize}

\subsubsection{Scale calibration}
The differential variances are obtained in unit of pixel square and
require a calibration of the pixel size. This was done by making
image sequences of the star $\alpha$ Centauri. {\bf It is an orbital
bright binary star whose angular separation (around 10 \arcsec) was
computed from its last orbit (Pourbaix \etal \cite{pourbaix})}.

Average autocorrelation of the images of the binary were computed to
reduce noise (one image sequence is around 600 images). This kind of
processing is well known in speckle interferometry to measure double
star separation. This function exhibits 3 peaks whose distance is
the separation of the binary stars in pixels. This gave a pixel
scale of $\xi=0.684\pm 0.004$~\arcsec (with binning $2\times 2$).

\subsubsection{Strehl ratio of DIMM images}
The Strehl ratio is an estimator of the quality of the two stellar
images produced by the DIMM. It is the ratio of the star's image
intensity at its maximum to the intensity of the theoretical Airy
disc that would have been obtained in perfect conditions. The Strehl
ratio is affected by fixed aberrations as well as optical
turbulence. It is generally assumed that image quality is good when
the Strehl ratio is over 30\%.

Monitoring the Strehl ratio of the two stellar images produced by a
DIMM can provide an image selection criterion. A simple calculation
formula has been proposed by Tokovinin (\cite{toko}). Though
continuous monitoring of the Strehl is not implemented in the data
acquisition software, we performed an {\em a posteriori} estimation
of the Strehl ratio of our DIMM images in typical conditions. From
data taken in the 6 days period of 10-15 December 2004, we estimated
the Strehl ratio of the two stellar images for each short-exposure
frame. We collected around $3\, 400\, 000$ values {\bf and found
average Strehl ratios $\langle S_l \rangle = 0.56 \pm 0.11$ for
image on the left and $\langle S_r \rangle = 0.53 \pm 0.11$ for the
one on the right.} These values indicate good image quality. Indeed,
Airy rings around the twin images were indeed often observed at the
DIMM's eyepiece.

\subsection{Isoplanatic angle measurements}
The isoplanatic angle was monitored during the month of January
2004. As for the DIMM, the telescope used for monitoring the
isoplanatic angle is a Celestron C11. A mask with a 10~cm aperture
and 4~cm central obstruction was placed at the entrance pupil. {\bf
Monitoring was performed from Jan 5 to Feb 2, 2004}.

The observing procedure was similar to the DIMM. The same star
Canopus was used. Exposure times from $\tau=8$ to 12~ms were used.
To compensate from exposure bias, we alternated frames with exposure
times $\tau$ and $\tau/2$. Time series were divided into 2~mn
intervals. Each short-exposure frame was applied the following
operations:
\begin{itemize}
\item Background mean level $\bar{b}$ was estimated on the whole image then subtracted
\item Low level values were set to zeros. Threshold was chosen at $5 \sigma_b$, $\sigma_b$ being the background variance
\item After these operations, star image was spreads over $N_I\simeq 100$ pixels in $2\times 2$ binning mode and $N_I\simeq 250$ pixels without binning. Total stellar flux $I$ was estimated by integration over these illuminated pixels.
\item Values of $I$, $\bar{b}$,  $\sigma_b$ and $N_I$ were logged in a file
\end{itemize}

A 2~minutes sequence corresponds to $N\simeq 3300$ images in $2\times 2$ binning mode, and to $N\simeq 1400$ images without binning. One sequence leaded to one value of the isoplanatic angle, computed as post-processing following the algorithm described hereafter:
\begin{itemize}
\item Separation of the values corresponding to exposure times $\tau$ and $\tau/2$ in two subsets
\item On each subset, computation of $\bar{I}$, $\sigma_I$ and scintillation indexes $s_\tau$ and $s_{\tau/2}$.
\item Compensation from exposure time {\bf by linear extrapolation} (Ziad \etal \cite{ziad})
\item {\bf Calculation of  $\theta_0$ for $\lambda=0.5 \: \mu$m.}
\end{itemize}

\section{Results}
\subsection{Seeing monitoring}
\begin{figure}
\includegraphics[width=\figwidth]{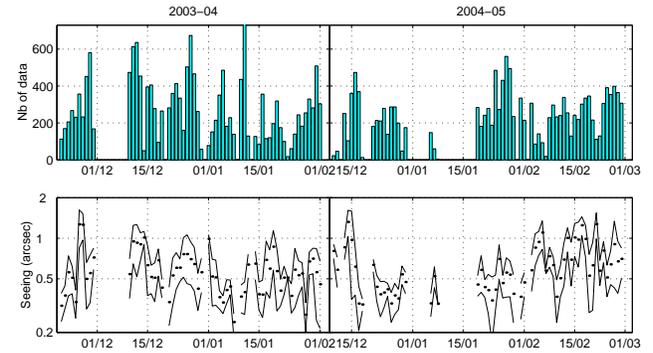}
\caption{Top: Number of seeing data per day. Bottom: points are
daily median seeing values  and interval containing 50\% of values
is delimited by lines. Data collected during the last two summer
campaigns. Seeing axis is logarithmic} \label{fig:seeingvsday}
\end{figure}

A total amount of 31597 2-minute seeing values have been estimated
during the campaigns 2003-2004 and 2004-2005. Thanks to the presence
of two observers, several long time series have been made possible,
making the monitoring as continuous as possible. As the polar
alignment was progressively improved, we could then leave the
telescope alone almost 8 hours without losing the star.
Figure~\ref{fig:seeingvsday} (top) shows the number of seeing values
obtained every day during the two campaigns (maximum possible value
is 720). Several periods of lack of measurements were due to bad
weather (as in December 2003 where we had 10 successive days of
covered sky) or logistics (this is the case in mid-January 2005
where part of the equipment was transferred into the Concordia
buildings and some work was done to set up the remote control).

\begin{figure}
\includegraphics[width=\figwidth]{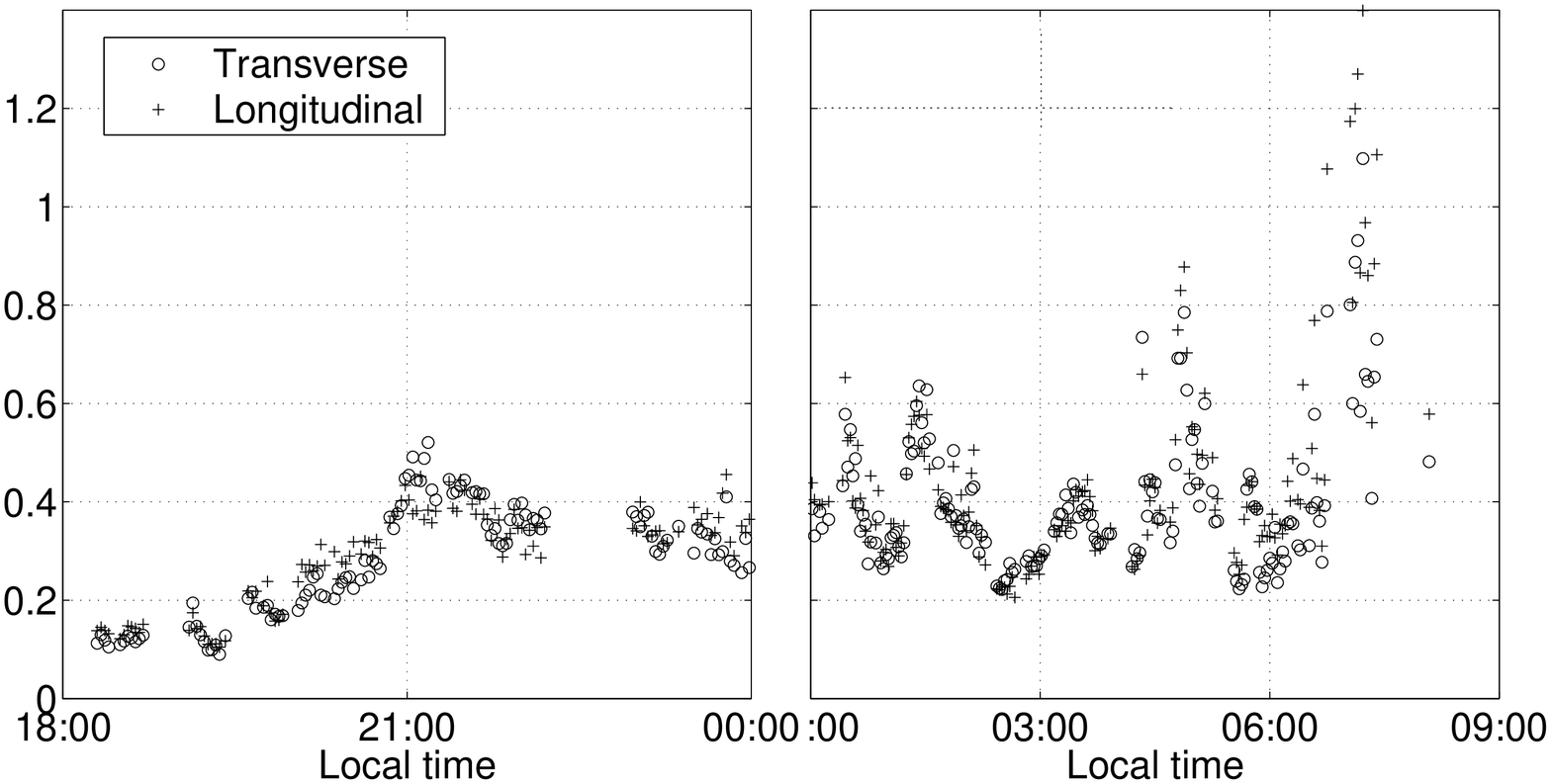}
\caption{First seeing curves obtained during the campaign 2003-2004 on November 21 and 22, 2003. We show here the longitudinal (+) and transverse (o) time series.}
\label{fig:seeing212111}
\end{figure}

Amazingly low seeing values were observed during the first days of
the 2003-04 campaign. November 21 corresponds to spring in southern
hemisphere; temperature was then close to -50$^\circ$C. These
temperature conditions are the closest to winter values we had on
the 3 campaigns. First 2 days seeing times series are shown on
Fig.~\ref{fig:seeing212111}. Exceptional seeing as low as
0.1\arcsec\ has been observed, and we had a continuous period of
10~hours of seeing below 0.6\arcsec. Daily median values are plotted
on Fig.~\ref{fig:seeingvsday}.

Seeing statistics are summarized in Table~\ref{table:seeingstat}. As
mentioned above, all measurements are computed at $\lambda=500$~nm
in daytime. Seeing values are in arcsec.

\begin{figure}
\includegraphics[width=\figwidth]{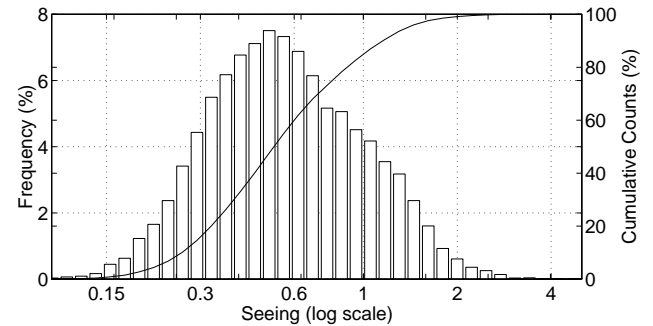}
\caption{Histogram (stairs) and cumulative histogram (continuous line) of seeing values for the campaigns 2003-2004 and 2004-2005. Seeing axis is in logarithmic scale.}
\label{fig:histoseeing04}
\end{figure}

\begin{table}
\caption{\bf Seeing statistics for the two summer campaigns. These number stand for the DIMM at $h=8.5$~m.}
\begin{center}
\begin{tabular}{l|r|r|r}\hline
Campaign                             & 2003-04 & 2004-05 & Total\\ \hline
Number of measurements & 17128   & 14469    & 31597\\
Median seeing  (\arcsec)         & 0.54    & 0.55    & 0.55  \\
Mean seeing    (\arcsec)         & 0.65    & 0.67    & 0.66\\
Standard deviation   (\arcsec)   & 0.39    & 0.38    & 0.39\\
Max seeing        (\arcsec)      & 5.22    & 3.33    & 5.22\\
Min seeing       (\arcsec)       & 0.10    & 0.08     & 0.08\\ \hline
\end{tabular}
\end{center}
\label{table:seeingstat}
\end{table}
Seeing histograms are displayed on Fig.~\ref{fig:histoseeing04}.
Scale has been set to logarithmic (base 10) on the seeing axis to
emphasize the log-normal distribution of the values. The 50\%
percentile (the median) is at $\epsilon=0.55$~\arcsec: seeing is
then better than 0.55 \arcsec half of the time.

These values are exceptionally good for daytime seeing, when the Sun
is present in the sky and heats the surface. It can compare with
night-time seeing of the best observatories.
Table~\ref{table:compseeing} shows a comparison with other sites.

\begin{table*}
\caption{Comparison of Dome C daytime seeing with daytime and night-time seeing in other observatories.}
\begin{tabular}{l|l|l|l|l|l}
\multicolumn{3}{l|}{Daytime values}     &  \multicolumn{3}{l}{Night-time values} \\ \hline
Site & Seeing & reference               & Site & Seeing & reference  \\ \hline
White Sands & 2.24 &    Walters et al. \cite{walters} & Paranal &   0.66 &  Sarazin  (www.eso.org/$\sim$msarazin)\\
Sac Peak &  1.68 &  Ricort et al. \cite{ricort}& Mauna Kea & 0.63 & Tokovinin \etal  \cite{tokoaz}\\
Roque de Muchachos   & 1.91 &   Borgnino \& Brandt \cite{borgnino}& Roque de Muchachos & 0.64 & Mu\~noz \etal \cite{munoz}\\
Sac Peak &  1.16 & Brandt et al \cite{brandt} & Cerro Pachon & 0.89 & Ziad \etal \cite{ziad} \\
Calern &    2.5 &   Irbah et al. \cite{} 1994& Maidanak  & 0.69 &   Eghamberdiev \etal \cite{ehgamberdiev} \\
Fuxian Lake & 1.20 &    Beckers \& Zhong \cite{irbah}& South Pole   & 1.74 &    Travouillon et al \cite{trav} \\\hline
Dome C  & 0.54 &    this paper & Dome C (day)   & 0.54 &    this paper \\ \hline
\end{tabular}
\label{table:compseeing}
\end{table*}

\begin{figure}
\includegraphics[width=40mm]{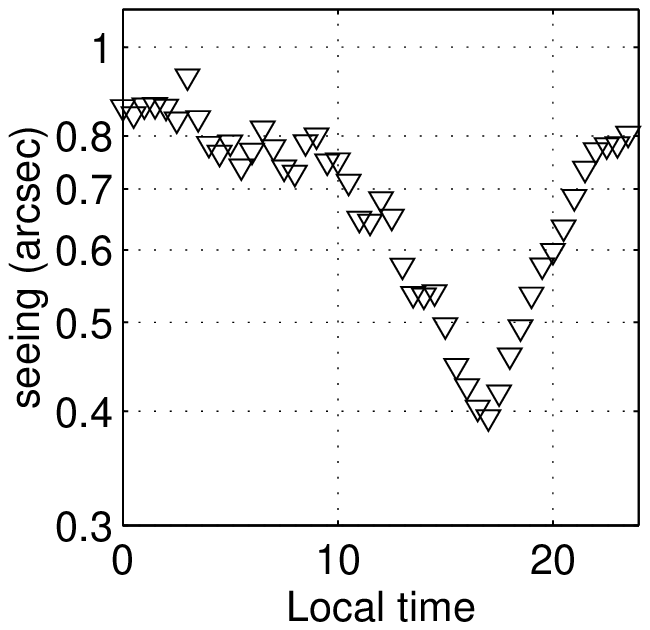}
\includegraphics[width=40mm]{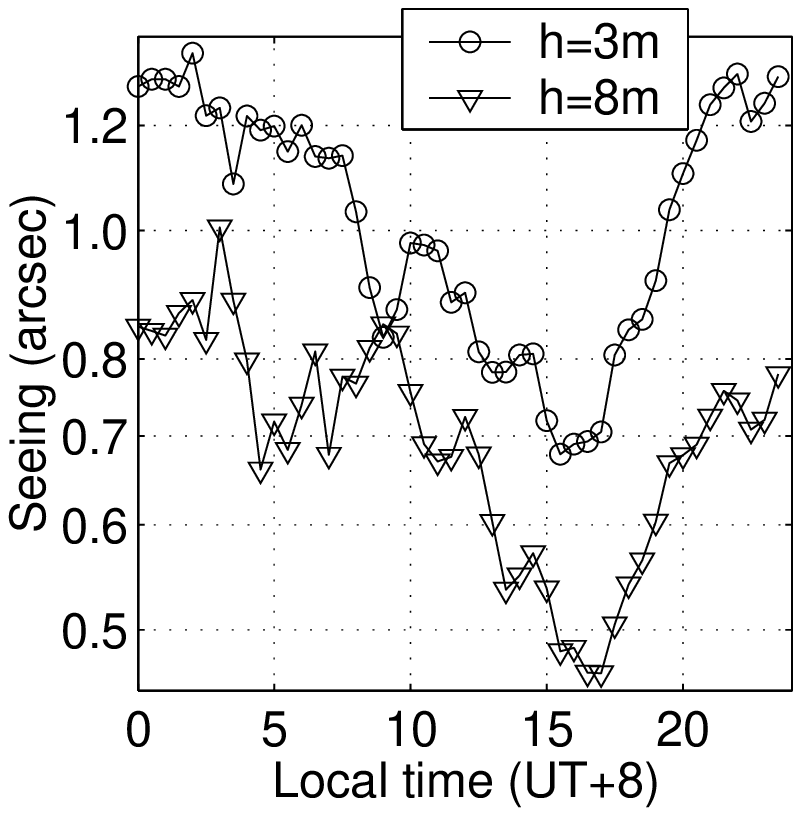}
\caption{Left: seeing versus time, averaged over the campaigns 2003-04 and 2004-05. Seeing values, obtained from the DIMM at elevation $h=8$~m, have been binned into 30~min intervals. Right: seeing versus time in 2004-2005 for the two DIMMs.}
\label{fig:seeingvstime}
\end{figure}

Another interesting result is the behaviour of the seeing with time.
Figure~\ref{fig:seeingvstime} has been calculated by binning all
seeing values into 30~min intervals. Best seeing values, like 0.4
\arcsec or better, are generally obtained in mid local afternoon. It
is extremely encouraging for solar imaging at high angular
resolution. Indeed, a discontinuity of the temperature gradient
between 200 and 400~m has often been noted in the middle of the day,
and disappears in the evening to be replaced by  a standard surface
inversion layer of 20 or 30~m (Aristidi et al. \cite{windpaper}).
During the afternoon transition, there is a moment with an
isothermal temperature profile. The generally excellent seeing
obtained during this transition indicates that the contribution of
all the rest of the atmosphere is indeed very small. At night with a
telescope standing above the ground inversion layer, a really
excellent seeing could then be expected almost continuously. The
height of this inversion layer is an open question that will be
answered after the winterover.

\subsubsection{Contribution of the surface layer}
During the 2004-05 summer campaign, the presence of two DIMMs
observing simultaneously at two different heights (3~m and 8~m over
the plateau snow surface) allowed an investigation of the
contribution of the surface layer to the seeing. Radiosoundings had
already shown that the strongest thermal gradients are observed
close to the surface (Aristidi \etal \cite{windpaper}) and an
important part of the turbulence is expected to be generated in the
first tens of meters.

All the telescopes were installed in the first days of December
2004, and the seeing monitoring started for telescope 1 (on the
ground) and 3 (on the platform) on December 10. Median/mean seeings
are 0.55/0.67 \arcsec for telescope at 8~m and 0.93/1.03 \arcsec for
telescope at 3~m (statistics over 15\,000 values obtained in
December 2004, January and February 2005). There is an important
difference which appears to be time-dependent. Figure
\ref{fig:seeingvstime} shows the behaviour of the seeing measured at
the two heights, as a function  of local time. Both curves exhibit a
noticeable minimum in mid-afternoon, though less pronounced for the
3~m curve.

We can describe the surface layer contribution with a turbulent
energy ratio (TER), following Martin \etal (\cite{martin}): \be
TER=\frac{\int_{\mbox{\scriptsize 3m}}^{\mbox{\scriptsize 8m}}
C_n^2(h)\, dh}{\int_{\mbox{\scriptsize 3m}}^\infty C_n^2(h)\, dh}
\ee This $TER$ gives the ratio of the turbulent energy in the 5~m
surface layer to the total turbulent energy (integrated from 3~m to
infinity). These integral can be estimated from the seeing through
the Fried's parameter (Roddier \cite{roddier}). The $TER$ is then
given by \be TER=\frac{r_0(3 m)^{-5/3}-r_0(8 m)^{-5/3}}{r_0(3
m)^{-5/3}} \ee
\begin{figure}
\includegraphics[width=\figwidth]{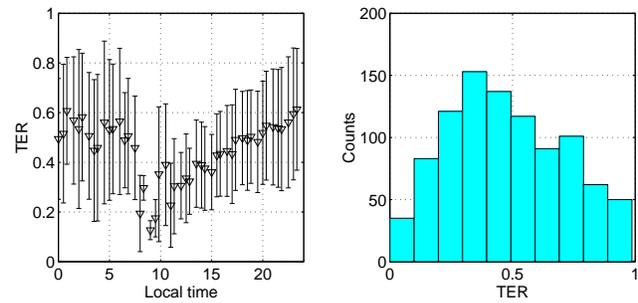}
\caption{Left : surface layer turbulent energy ratio ($TER$) as a function of time. Error bars are the standard deviation of the samples distribution. Right : $TER$ histogram.}
\label{fig:ter}
\end{figure}
The $TER$ was calculated every time the two telescopes were operated
simultaneously. Its histogram and time-dependence are shown on
Fig.~\ref{fig:ter}. Mean value of 48\% indicates that almost half of
the ground turbulence is concentrated into the first 5~m above the
surface.

\subsection{Isoplanatic angle and scintillation}

More than 6000 values of the isoplanatic angle have been collected
during the month of January, 2004. Statistics of both $\theta_0$ and
the scintillation index $s$ summarised in Table~\ref{table:statisop}
show a median value $\theta_0=6.8$~\arcsec at wavelength
$\lambda=0.5 \mu$m, which is far better than values obtained in any
astronomical site (see Table~\ref{table:compisop} for a comparison).
Good values are found also at the South Pole where the atmosphere
above the first 220~m is calm (Marks \etal \cite{marks}); the
isoplanatic angle is indeed more sensitive to high  altitude
turbulence. Winter estimates given by the MASS in the AASTINO
station (Lawrence \etal \cite{nature}) indicates values similar to
ours, despite the presence in winter of high-altitude winds of the
order of 40~m/s (Aristidi \etal \cite{windpaper} and references
therein).

\begin{figure}
\includegraphics[width=\figwidth]{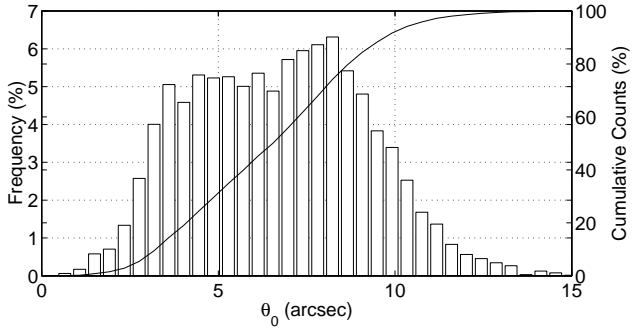}
\caption{Histogram (stairs) and cumulative histogram (line) of isoplanatic angle values for the campaign 2003-2004.}
\label{fig:histoisop04}
\end{figure}

\begin{figure}
\includegraphics[width=\figwidth]{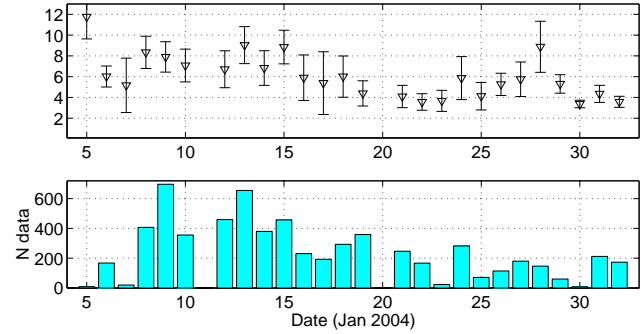}
\caption{Isoplanatic angle as a function of day. Top: daily median values. Error bars corresponds to the $1\sigma$ dispersion of the daily values. Bottom: number of data per day.}
\label{fig:isopvsday04}
\end{figure}

The histogram and cumulative histogram are shown in
Fig.~\ref{fig:histoisop04}, time series on
figure~\ref{fig:isopvsday04}. The daily median values plot in the
top of Fig.~\ref{fig:isopvsday04} shows a small degradation of
$\bar\theta_0$ between the beginning and the end of January.

The large value of isoplanatic angle, roughly 3 times larger than in
classical sites, is a good news for adaptive optics. This
corresponds to a gain of a factor 10 in the field usable to find
calibrator stars, and therefore increases the observable piece of
sky, as discussed by Coud\'e du Foresto (Coud\'e du Foresto \etal
\cite{vcdf}). Another advantage of a large isoplanatic domain is the
uselessness of multi-conjugate adaptive optics for high-resolution
wide field imaging (Lawrence \cite{lawrence}).

\begin{table}
\caption{Isoplanatic angle ($\theta_0$) and scintillation index statistics during the month of January 2004.}
\begin{center}
\begin{tabular}{l|r|r}
                       & $\theta_0$ (\arcsec) & $s (\%)$ \\ \hline
\# of measurements & 6368 & 6368\\
Mean value             & 6.8 & 0.88\\
Median value           & 6.8 & 0.63\\
standard deviation     & 2.4 & 0.90\\
Min value              & 0.7 & \\
Max value              & 17.1 & \\ \hline
\end{tabular}
\end{center}
\label{table:statisop}
\end{table}

\begin{table}
\caption{Comparison of Dome C isoplanatic angle with values observed in other sites.}
\begin{tabular}{l|l|l}
Site & $\theta_0$ & reference  \\ \hline
Paranal & 1.91 &    Ziad \etal  \cite{ziad} \\
La Silla & 1.25  &  Ziad \etal  \cite{ziad} \\
Cerro Pachon (Chile) & 2.71 &   Ziad \etal  \cite{ziad} \\
Maidanak & 2.47 &   Ziad \etal  \cite{ziad} \\
Oukaimeden (Morocco) & 1.58 &   Ziad \etal  \cite{ziad} \\
South Pole & 3.23 & Marks \etal \cite{marks} \\
Dome C  & 6.8 & this paper \\ \hline
\end{tabular}
\label{table:compisop}
\end{table}


\section{Discussion and conclusion}
We have presented the results of optical turbulence measurements
during two summer campaigns at Dome C. The main result is the
exceptional seeing quality in the daytime, allowing image resolution
better than 0.5 \arcsec during a few hours every day, and the large
value of the isoplanatic angle, three times larger than Mt Paranal
in night-time. Combining this with large periods of clear and
coronal sky makes Dome C probably one of the best sites on earth for
solar visible and infrared astronomy.

\begin{figure}
\includegraphics[width=75mm]{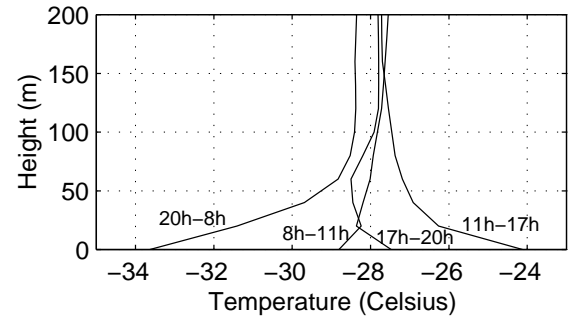}
\caption{Mean temperature profiles above the ground (from Aristidi \etal \cite{windpaper}), based on in-situ radiosoundings. On the vertical axis, height is counted from the snow (altitude 3260m). The four curves correspond to measurements performed at four different times of the day}
\label{fig:polytemp}
\end{figure}

We recently published (Aristidi \etal \cite{windpaper}) a study
based on balloon-borne meteo radio-sondes launched during 4 summer
seasons, allowing us to make statistics on wind speed and
temperature profiles in the atmosphere above Dome C. Among the
numerous results presented in that paper, we found that the
temperature profile exhibit strong gradient in the boundary layer
(the first 100~m above the snow). Corresponding curve is shown in
Fig.~\ref{fig:polytemp}. This gradient, positive at midnight(ice is
cooler that the air above) and negative at noon (ice is heated by
the Sun radiation), vanishes twice a day: in the morning and near
5pm. Seeing appears indeed to be the best during the afternoon near
5pm. {\bf The other expected seeing minimum in the morning has been
sometimes observed, especially in the 2003-2004 campaign. But it
does not appear in the daily averaged curves displayed in
Fig.~\ref{fig:seeingvstime} and at this time we have no convincing
explanation for that.}

\begin{figure}
\includegraphics[width=\figwidth]{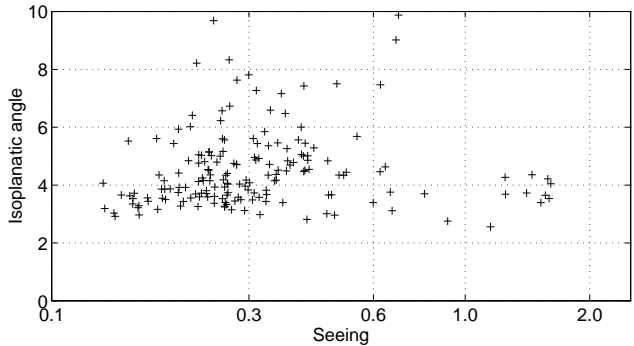}
\caption{Plot of the isoplanatic angle versus seeing (data collected in January 2004).}
\label{fig:seeingvsisop}
\end{figure}

This behaviour of the boundary layer temperature profile and of the
seeing-versus-time curve suggests that the turbulence is dominated
by the first tens of meters above the ground. This is the same at
the South Pole, where the height of this very turbulent boundary
layer is 220~m (Marks \etal \cite{marks}). Indeed, another indicator
is the correlation between the seeing and the isoplanatic angle.
Both result from an integral over the whole atmosphere, but the
isoplanatic angle is more sensitive to high turbulent layers
(ponderation by $h^{5/3}$ in the integral definition of $\theta_0$
(Ziad \etal \cite{ziad})). Figure~\ref{fig:seeingvsisop} displays a
plot of the  isoplanatic angle versus seeing showing no dependence
between the two parameters. Quantitative estimations of the surface
layer contribution have been made possible in 2004-2005 with the
presence of two DIMMs. 50\% of the ground seeing is generated in the
first 5~m.

Now, the winter measurements are awaited with a lot of excitement,
to know how much residual turbulence will exist above this ground
layer during the winter night, how thick will be this layer, and how
much turbulence will exist below. The last summer campaign was the
first one to be immediately followed by the historical first
winter-over, and Agabi has volunteered to spend one year at
Concordia to conduct the observations. Seeing and isoplanatic angle
monitoring are in progress. In-situ soundings of the vertical
profile of $C_n^2$ by means of balloon borne microthermal sensors
are also at the menu for the winter. They will give access to
parameters such as the outer scale and coherence time. Finally a
monitoring of $C_n^2$ in the boundary layer is also foreseen, using
a ground version of the balloon experiment which takes advantage of
the 32~m high American tower.

At the time of writing this paper, the night is getting longer and
longer every day and these questions will receive firm answers quite
soon.

\begin{acknowledgements}
We wish to acknowledge both Polar Institutes, IPEV and ENEA for funding the program and for the logistics support. Thanks to all the Concordia staff at Dome C for their friendly and efficient help in setting up the Concordiastro platforms. We are also grateful to our industrial partners ``Optique et Vision'' and ``Astro-Physics'' for the technical improvement done on the telescopes and their mounts to make them work in polar conditions. Finally we thank Jean-Michel Clausse and Jean-Louis Dubourg who were present on site during the campaings 2000-2001 and 2001-2002.
\end{acknowledgements}

\appendix
\section{Error analysis}
\subsection{Seeing}
\paragraph{Statistical error.}
Variance of image motion is computed from a sample of $N\simeq 9000$
individual frames: it is then affected by statistical noise due to
the finite size of the sample. Assuming statistical independence
between the frames, the statistical error on the variance $\sigma^2$
is given by (Frieden, 1983, Sarazin \& Roddier, 1990) \be
\frac{\delta_s \sigma^2}{\sigma^2}=\sqrt{\frac{2}{N-1}} \ee that
propagates onto the seeing an error contribution $\delta_\epsilon$.
With 9000 independent frames we have $\frac{\delta_s
\sigma^2}{\sigma^2}=1.4$\% and
$\frac{\delta_s\epsilon}{\epsilon}=0.9$\%. Frames are not
independent at our sampling rate and this is only a lower boundary.
\paragraph{Scale error.}
Image motion is converted from pixels to arcsec using the factor
$\xi$ introduced before. The uncertainty on $\xi$ propagates into
the differential variances when the conversion from pixels into
arcsec is performed. {\bf With actual value
$\frac{\delta\xi}{\xi}=0.006$ that gives a relative contribution on
the differential variances $\frac{\delta_p
\sigma^2}{\sigma^2}=1.2$\% and on the seeing
$\frac{\delta_p\epsilon}{\epsilon}=0.7$\%}.

\paragraph{Readout noise}
Influence of the CCD readout noise on DIMM data is discussed in
Tokovinin, 2002. The readout noise is a random independent
contribution to the measured flux. It biases the computed
differential variances by a term \be \sigma_R^2=2
\frac{R^2}{I^2}\sum_{\mbox{\scriptsize window}} x^2_{ij}
\label{eq:ron} \ee where $I$ is the total stellar flux, $R$ is the
readout noise (2.3 ADU for the Pixelfly) and $x_{ij}$ the
coordinates of contributing pixels (the number of illuminated pixels
is of the order of 30 after flat fielding and that defines the
``window'' over which the summation is made). The order of magnitude
of this bias in our case is $\sigma_R^2\simeq 10^{-6}$ square
pixels. Comparing this value to our standard differential variances
(0.1 to 1 square pixel), we can see that the readout noise bias is
extremely small and can be neglected.

\paragraph{Background noise.}
The sky background is an additive Poisson noise independent from the
stellar signal. Therefore its effects on the differential centroid
variance is the same as the readout noise: a bias term $\sigma_B^2$.
It can be computed using eq.~\ref{eq:ron}, substituting $R$ by $B$,
the background standard deviation (square root of background flux
per pixel). The background is a function of time, as shown by
Fig.~\ref{fig:levelvstime}; it can attain 30\% of the stellar flux
when the Sun is at its maximum (typical values in ADU for the
highest background are $B\simeq 1000$). It leads to a bias term
$\sigma_B^2\simeq 10^{-4}$ square pixels which is still negligible
compared to the differential variance values.
\subsection{Isoplanatic angle}

\paragraph{Background Noise.}
The presence of a strong background on individual images causes
uncertainties and biases on the estimation of the mean stellar
intensity $\bar{I}$ (ensemble average over the image sample), its
standard deviation $\sigma_I$ and then on the scintillation index
$s$. As shown on Fig.~\ref{fig:levelvstime}, the background can be
as high as 30\% of the stellar flux when the Sun is at its maximum.
To perform a bias an SNR estimation, let us introduce the following
variables:
\begin{itemize}
\item $B$, the background intensity collected over the $N_I$ pixels illuminated by the star after threshold application, $\bar{B}$ and $\sigma^2_B$ its mean and variance. $B$ is a Poisson random variable, it must verify $\sigma_B=\sqrt{B}$, that was well verified on images.
\item $I_t$ the total intensity (background+stellar flux) collected over the $N_I$ pixels.
\end{itemize}
The stellar flux is given by $I=I_t-B$, the measure being $I_t$. The
mean $\bar{I}$ is biased by the term $\bar{B}$. This bias is
estimated (we assume stationarity so that the ensemble average
$\bar{B}$ is equal to the average over one image) and removed as
indicated above, but the background fluctuations lead to an error
$\delta I$ on the estimation of $\bar{I}$ equal to $\delta
I=\sigma_B\simeq \sqrt{B}$.

The variance $\sigma_I^2$ is equal to the difference
$\sigma_I^2=\sigma_{It}^2-\sigma_B^2$ assuming independence between
the stellar flux and the background. The variance estimation we make
on images is $\sigma_{It}^2$, it is then biased by the term
$\sigma_B^2$. However we remarked that this bias is less than 1\% of
$\sigma_{It}^2$ and decided not to debias the variances. In addition
to this bias, there is an error term $\delta \sigma_{B}^2$ due to
the uncertainty of the estimation of $\sigma_B^2$. Hence the total
error on $\sigma_I^2$ is $\sigma_B^2+\delta \sigma_{B}^2$ if we do
not debias the variances.

The scintillation index is the ratio $s=\sigma_I^2/\bar{I}^2$; its
error $\delta s$ can be estimated by \be \frac{\delta
s}{s}=\frac{\delta \sigma_I^2}{\sigma_I^2}+2 \frac{\delta
\bar{I}}{\bar{I}}= \frac{\sigma_B^2+\delta
\sigma_{B}^2}{\sigma_I^2}+2\frac{\sigma_B}{\bar{I}}
\label{eq:bcnoise}\ee Typical values corresponding to the worst case
(strongest background at Sun's maximum elevation) are, in ADU units:
$\sigma_B\simeq 400$, $\bar{I}\simeq 40000$, $\sigma_I\simeq 5000$
and $\delta\sigma_B^2\simeq 70000$. That gives a background error
contribution $\frac{\delta s}{s}=3\%$

\paragraph{Readout noise.}
The readout noise can be considered as a Gaussian random variable
with mean $r$ (per pixel) and standard deviation $\sigma_r=2.3$~ADU
(from Pixelfly documentation). As the star is spread over $N_I$
pixels, we will consider the variables $R= N_I r$ (mean over the
$N_I$ pixels) and its standard deviation $\sigma_R=\sqrt{N_I}
\sigma_r$. The mean value $R$ is automatically removed by the
background substraction.  The same reasoning as above can be applied
to the readout noise. From Eq.~\ref{eq:bcnoise} the contribution
$\frac{\delta_r s}{s}$ of the readout noise is then given by \be
\frac{\delta_r s}{s}\simeq
\frac{\sigma_R^2}{\sigma_I^2}+2\frac{\sigma_R}{\bar{I}} \ee in which
we have neglected the term $\delta \sigma_R^2$. Taking the same
values than for the background noise we have $\frac{\delta_p
s}{s}\simeq 10^{-3}$ that is one order of magnitude below the
background noise and can be neglected.

\end{document}